\documentclass[10pt, conference, compsocconf]{IEEEtran}

\usepackage{graphicx}

\ifCLASSINFOpdf
\else
\fi
\usepackage{makeidx}
\usepackage{amsmath}
\usepackage{epsfig}
\usepackage{amssymb}
\usepackage{textcomp}
\usepackage{url}
\usepackage{pstricks,pst-text,psfrag}

\usepackage{mathtools}

\usepackage{algpseudocode}
\usepackage{algorithm}

\usepackage{subfig}

\usepackage{url}

\begin{document}
%
\title{\LARGE \mbox{LWB and FS-LWB implementation for Sky platform using Contiki}}

\author{\IEEEauthorblockN{Chayan Sarkar}
\IEEEauthorblockA{Delft University of Technology, The Netherlands\\
Email: chayan@ieee.org}}
\maketitle

%

\begin{abstract}
The low-power wireless bus (LWB) is a highly efficient communication protocol for wireless sensor networks (WSN). However, a lack of open implementation for this protocol prompted the implementation of the protocol for the most common platform of WSN related research, the `sky' nodes using the Contiki operating system. This document provides the detailed description of the implementation.

A forwarder-selection mechanism is developed to improve energy-efficiency of LWB in data collection scenario. The implementation also consists the forwarder selection mechanism. One can either use the original features of LWB or enforce the forwarder selection mechanism.
\end{abstract}


%
\IEEEpeerreviewmaketitle


\section{Introduction}
A wireless sensor network (WSN) consists of a number of embedded sensor devices, where the nodes\footnote{A sensor device and a node are synonymous.} collect data using their sensors and report it to a central collection point, called the sink node. If the sink is connected to the Internet, the data can be accessed over the Internet, i.e., the WSN deployment region can be monitored remotely. The ease of deployment and management broadened the use of WSN and makes it a very useful tool for the Internet of Things (IoT).

Most of the WSN devices are battery operated with the sink node as a possible exception. Thus, WSN related research is primarily focused on energy-efficient network operations as energy is one of the most precious resources. As a result, the routing protocols for other types of ad-hoc networks are not suitable for WSN. The routing protocols for WSN are primarily focused towards reducing the duty cycle (radio on time) of the nodes.

Over the years, many routing protocols have been proposed and developed. Recently, the low-power wireless bus (LWB)~\cite{ferrari2012low} has been proposed by Ferrari \textit{et al.}. Using extensive experiments, it is shown to be the most energy efficient communication mechanism till its inception. The design of LWB is very generic and simple. Also, there is a drastic difference in its design as compared to traditional communication mechanism in WSN. This makes it a very attractive choice as a communication protocol. However, there is a lack of open implementation of LWB. This restricts the usage and further development of protocols in the academia and industry. This article provides an open implementation of the protocol for Tmote Sky platform. The reason behind choosing the Sky platform is that it is one of the most common sensor node platform among in the WSN research community. Most of the online testbed (e.g., Indriya~\cite{doddavenkatappa2012indriya}, Flocklab~\cite{lim2013flocklab}) also consists of Sky-based devices. The goal of this article is to enable development of communication mechanisms based on LWB as well as provide a comparison platform for the new protocols.

As mentioned earlier, LWB is a very generic protocol and it can cater a vast range of application scenarios. This also provides the scope of improvement for a particular use-case. Forwarder selection low-power wireless bus (FS-LWB)~\cite{forwarderdcoss2013} is one such optimization of LWB for data collection applications. This implementation also provides a plug-in to enable forwarder selection on top of LWB. The code is available in~\cite{lwb_code}.

\section{Background }
Before providing the implementation details, we describe the basic concept of LWB and the optimization achieved by FS-LWB. However, we highly recommend to read the original articles by Ferrari \textit{et al.}~\cite{ferrari2012low} and Carlsen \textit{et al.}\cite{forwarderdcoss2013}, respectively. 

The development of LWB stands on the shoulder of a fast and efficient network-wide flooding mechanism, called Glossy~\cite{ferrari2011efficient}. In Glossy, the master clock holder (the sink node) sends a periodic sync packet. Every node, after receiving the sync packet, rectifies its clock offset and becomes synchronized with the sink node. In contrast with traditional clock synchronization  protocol where the process of resynchronization takes a long time, glossy can quickly resynchronize the whole network. This fast synchronization is an effect of the fast flooding mechanism. In glossy, nodes that are far away from the sync, they too receive the sync packet within a small time-bound. This delay is compensated based on hop distance of a node from the sink. The sync packet contains a counter that is initiated to 1 by the sink. Every node after receiving the packet increases the counter before forwarding it. Thus, based on the counter value of the received packet, a node determines its hop distance from the sink. 

Glossy achieves an incredibly fast network-wide flooding by utilizing a phenomenon called constructive interference (CI). When multiple nodes transmit the same content simultaneously, the signals interfere constructively at the receiver. As a result, the receiver can successfully decode the content of the transmitted packet. In Glossy, every node wakes up (turns on the radio) just before the start of a new flood. The sink sends the sync packet and all its first hop neighbors receive its. Packet reception at a node triggers the packet transmission in by the node. Glossy ensures a fixed delay for receive-to-transmit switching delay at every node. Thus, when the first hop nodes transmit the packet (immediately or after a small but fixed delay), they transmit concurrently. This ensures CI at the second hop nodes. As a result, all the second-hop nodes are able to receive the packet at the same time. Then they also forward the packet immediately, which takes the packet further away from the sink. This ripple effect immediately covers the whole network and the flooding gets completed. As mentioned earlier, nodes implicitly become synchronized with the sink based on receiving the sync packet.

By utilizing Glossy as a communication primitive, the low-power wireless bus (LWB) protocol provides a fast and efficient communication protocol. In Glossy, the sink node is the only node that can start a flood, and the sync packet need not carry any useful payload. However, not only some payload can be attached to the packet to be flooded, the flooding can be started by any other node apart from the sink node. However, if two nodes start the flooding in an overlapping time, the flooding may fail partially or fully. This is because Glossy utilizes CI where one of the requirement is transmitting the same content by the simultaneous transmitters. Thus, the nodes need to coordinate their respective flooding time among themselves such that everyone floods at a non-overlapping time. Moreover, every node should know the number of flooding nodes in the network, and flooding time of each node. LWB arranges these by using a centralized scheduler (can be the sink node). LWB uses a slotted communication, where a slot length is the amount of time required to flood the whole network. Every node requests for a data slot from the sink (scheduler). In return, the scheduler assigns a unique slot for every node. 

Once the slot assignment is done, the data transmission is done using Glossy-based flooding, where every node initiates a Glossy flood in its assigned slot. LWB defines an LWB round. At the beginning of a round, the sink sends a sync packet, and the whole network becomes synchronized with the sink. Additionally, the sync packet contains the number of data slots that will follow in that LWB round. At the beginning of every slot, all node wakes up as a receiver except the node that is supposed to start the flood in that slot. Once a node performs it receive-and-forward operation in the flooding process, it goes back to sleep mode until the beginning of the next slot. Then the node either discards the packet or forwards the packet to the application layer depending on whether it is the intended recipient or not. As every communication is based on network-wide flooding, every packet is available to every node. Thus, LWB inherently supports all the traffic patterns, i.e., one-to-one, one-to-many, many-to-one, and all-to-all. One of the major benefits of LWB is that it does not require any control packet exchange to monitor the link quality or the network condition.

As every node participates in every data delivery slot in LWB, it unnecessarily wastes energy when every data packet need not be delivered to every node. For data collection scenarios, where data is only needed to be delivered to the sink, an optimization is proposed by Carlson \textit{et al.}~\cite{forwarderdcoss2013}. For every data source, it finds a set of forwarder nodes along the path from the source to the destination. If a node is on the way to the destination for or a particular source node, it participates in the corresponding flooding slot. Otherwise, it just avoids participating in that particular flooding slot. The forwarder set is discovered in a distributed way without any additional control overhead.

Similarly, another approach that improves the energy efficiency by dynamically adjusting the transmission power is described in~\cite{rao2016dipa}. However, this is out of the scope of this article. Thus, we do not discuss this article any further.

\begin{figure}
    \centering
    \includegraphics[width=\linewidth]{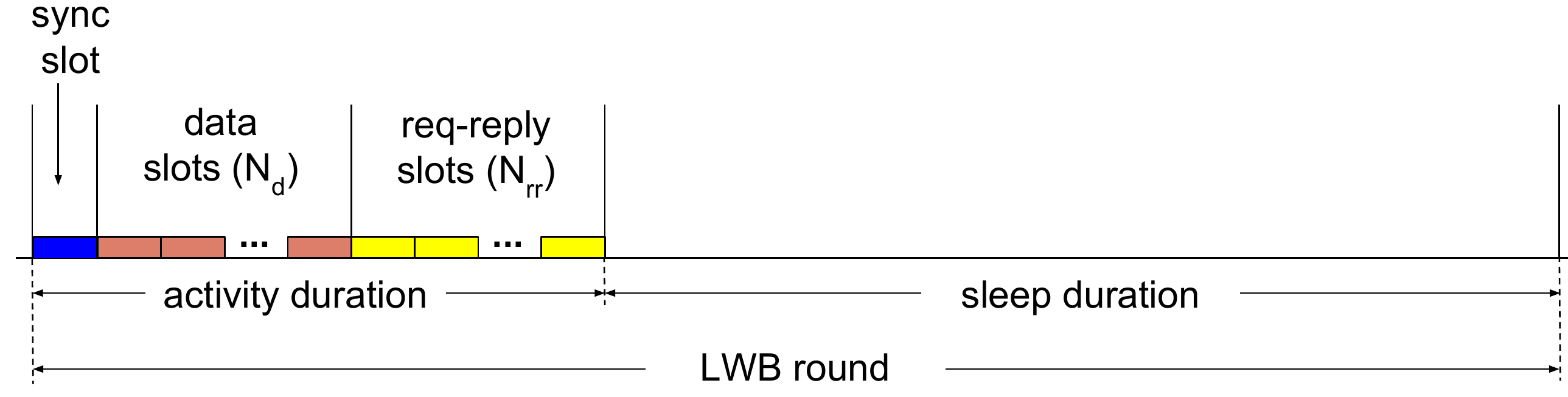}
    \caption{Slotted communication in LWB: showing different types of slots that are repeated in every LWB round.}
    \label{fig:lwb-round}
\end{figure}

\section{Implementation details}
This implementation is done using the Contiki operating system~\cite{dunkels2004contiki}. LWB uses synchronous and slot-based communication scheme. The slot structure of LWB is shown in Fig.~\ref{fig:lwb-round}. 

\subsection{LWB slot structure}
As mentioned earlier, every slot is equal to the length of a network-wide Glossy flood. There is three basic type of slots - (i) sync slot, (ii) request/reply slots, and (iii) data slots. Apart from the slots, LWB defines round, which is the periodicity of the sync packet. An LWB round starts with a sync packet, which is followed by other types of slots. However, it is not strict to have either the request/reply or the data slots in every LWB rounds. Also, the number of these slots varies among rounds. Details of each slot are described in the following. 

\subsubsection{Sync slot}
The sync slot is exclusively used by the sink node. By receiving this packet, every node becomes synchronized with the sink. The content of this packet indicates the number of request/reply and data slots in this LWB round.  Additionally, it also indicates the length of the LWB round, i.e., when the next sync packet will be sent. 

\begin{figure*}
    \centering
    \includegraphics[width=\linewidth]{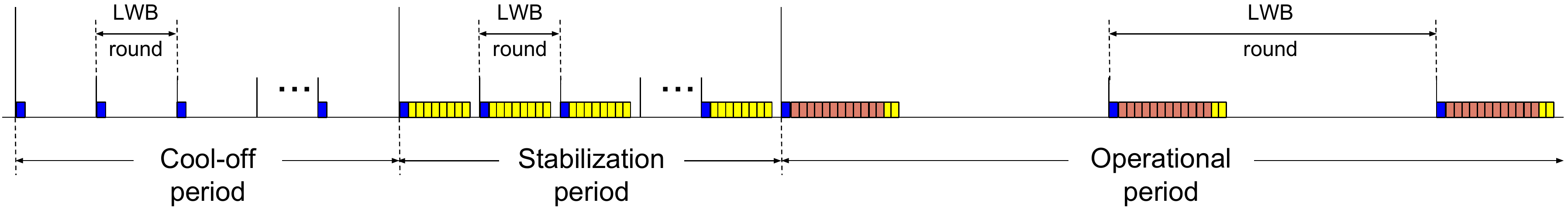}
    \caption{Working process of LWB from bootstrapping.}
    \label{fig:lwb-process}
\end{figure*}

\subsubsection{Request/reply (RR) slot}
These slots are used to obtain a unique data slots. In these slots, multiple nodes can contend to request a data slot from the sink. 
In the first slot, nodes contend to send a request. If sink receives a request successfully, it sends a reply in the second slot. The node who already acquires a data slot stops contending. In the third slots again the remaining nodes send a request. The process continues for the specified number of RR slots in an LWB round. If sink mentions $N_{rr}$ RR slots in an LWB round, there can be a maximum of $N_{rr}/2$ requests processed in that round. 

\subsubsection{Data slots}
After the intense request/reply period (see Section~\ref{sec:stabilization-period}), the sink node assumes that most of the nodes have acquired a data slot. Thus, the sink  node indicates the number of data slots (that it assigned so far), in an LWB round. In the successive LWB rounds, availability of a certain number of data slots ($N_d$) is indicated by the sync packet. In a particular data slot, the node who owns the slot, starts the flooding, while the remaining nodes just participate. 

\subsection{LWB operation in a long run}
The process of LWB operation is shown in Fig.~\ref{fig:lwb-process}. The whole process is divided into three phases/periods. The behavior of the process can be manipulated by changing the durations of these periods. Various parameters including the length of these periods can be customized in the specification file (slot-def.h). The three main phases of LWB operation are described below.

\subsubsection{Cool-off period}
The first step of LWB process is to get synchronized with the sync. As the nodes do not have any idea about the LWB slot structure, they can learn this only by receiving a sync packet. At the beginning, a node keeps its radio on to receive the first sync packet. During the cool-off period, the sink node sends a sync packet every second. During this period, an LWB round contains a sync slot only. Once the sync packet is received, the node knows the length of the LWB round, i.e., when the next sync packet will be sent. Thus, the node keeps it radio off until the beginning of the next round. This process is same as Glossy, except the content of the sync packet, which determines the time until the next sync packet and the number of other slots that may follow the sync packet. The idea behind cool-off period is to provide some time so that every node can join the network by synchronizing with the sink nodes.

\subsubsection{Stabilization period}
\label{sec:stabilization-period}
After some cooling off period (LWB round with only sync packets), the sink indicates a number of RR slots that will be followed. Please note, only the sink node knows the duration of the cool-off period. For other nodes, it does not matter, as they just perceive it as an LWB round with just a sync slot. 

In the stabilization period, the length of an LWB round is also fixed to 1 seconds. The sink indicates an availability of as many RR slots as it is possible to fit within one second. Hence, the nodes start requesting for a data slot. Please note at this point still there are no data slots in the LWB round. As the time progresses in the stabilization period, more nodes acquire a data slot and stops contending. Thus, if a high number of RR slots are mentioned, a lot of energy will be wasted to participate in the RR slot. We adopt a dynamic approach to adjust the number of RR slots in a round. 

Initially, it is set to the maximum that can fit in a second. As soon as the sink sees two request slots within any request, it reduces the number of RR slots to the minimum, i.e., 2 in LWB, and 3 in FS-LWB (see Section~\ref{sec:fs-lwb}). The authors of LWB described 2 RR slots in every LWB rounds. That means in every round only one node can acquire a data slot, and it takes a long time before every node acquires a data slot. In this implementation, data slot acquisition speeds up significantly.

\begin{table*}
	\renewcommand{\arraystretch}{1.3}
	\centering
	\caption{Table defining the application defined parameters (defined in file ``slot-def.h''). These parameters change the behavior of the LWB process and need to be adapted according to application requirements.}    
    \begin{tabular}{|r|l|l|}
        \hline
        \textbf{Parameter} & \textbf{Function} & \textbf{Default value} \\ \hline
        IPI & sensing frequency & 10\,s \\ \hline
        MINIMUM\_LWB\_ROUND & duration of minimum LWB rounds in operational period & 5\,s \\ \hline
        COOLOFF\_PERIOD & duration of the cool-off period & 10\,s \\ \hline
        STABILIZATION\_PERIOD & duration of the stabilization period & 10\,s \\ \hline
        MAX\_PAYLOAD\_LEN & defines maximum payload of an LWB data packet & 40 bytes \\ \hline
        SINK\_NODE\_ID & defines who is the sink node & 1 \\ \hline
        MAX\_NODE\_NUMBER & highest node-id should not cross this number & 150 \\ \hline
        FORWARDER\_SELECTION & enable forwarder selection & 0 (disabled) \\ \hline
    \end{tabular}%
	\label{table:params}
\end{table*}

\subsubsection{Operational period}
In the stabilization period, most of the nodes have acquired a data slot (if not all). In case there are some nodes who do not have a data slot yet or a node joins the network at a later stage, the minimum number of RR slots are maintained in every LWB round such that these nodes can acquire a data slot. Thus, during the operational period, an LWB round contains 1 sync packet, $N_d$ data slots (equals to the number of data sources in the network), and a minimum number of RR slots. 

The length of an LWB round is equal to the periodicity of the sensing application. If the sensing periodicity of the application is too high, the LWB round are cut into smaller length. This ensures that the nodes remain synchronized in between two round. Glossy can tolerate a certain clock drift by the nodes between the rounds, which is defined as the \textit{GLOSSY\_GUARD\_TIME}. It is set to 2\,ms. So the minimum length of an LWB round should be set in a way that the clock drift is contained within this bound. By increasing this guard time, a larger drift can be tolerated and a more infrequent LWB round is possible. However, larger guard time increase energy consumption by the nodes. If the minimum LWB round length is less than the data sensing frequency, the intermediate round would not contain any RR or data slots, but just a sync slot to keep the network synchronous. The relevant parameters that controls the behavior of the LWB process is summarized in Table~\ref{table:params} (defined in file ``slot-def.h'').

\subsection{Forwarder set determination}
\label{sec:fs-lwb}
In the case of traditional LWB, the RR slots are used in pairs, where odd numbered slots are used for requesting a data slot (via contention), and even numbered slots are used by the sink to reply with an assigned data slot. If the forwarder selection mechanism is enabled, the RR slots are used in a triplet. In that case, the every third slot is used by the source node who just acquired the data slot. The source node announces his hop-distance from the sink using this slot. The additional operations are as following.

\begin{itemize}
\item In the second RR slot, when the sink replies with the data slot assigned to the requester, every node calculates its hop distance from the sink along with forwarding the packet. Let's assume this distance is $h_u$.

\item Like any other nodes, the source itself also calculates $h_u$ after receiving the reply. 

\item In the third RR slot, the source announces its $h_u$ as $h$ along with the slot number $s$.

\item Upon receiving this announcement, every node calculates its hop distance from the source. Let's assume this distance is $h_d$.

\item After this, a node add slot $s$ to its forwarder set, if it is on the way from the source to the sink, i.e., $h = h_u + h_d$. Otherwise, during the operational phase, the node do not participate in data slot $s$.
\end{itemize}

\section{Conclusions}
The idea behind this document is to help developers to build and evaluate new routing protocols. The development can be an improvement over LWB or FS-LWB, or it can be a completely new approach. In both the cases, this implementation provides an easy way to compare the performance of the new protocols with respect to LWB and FS-LWB.

\section*{Ackowledgement}
This code is primarily based on Glossy~\cite{ferrari2011efficient} implementation and some parts from Chaos~\cite{landsiedel2013chaos}. I cordially thank the authors and the developers of these works. 


\tiny
\bibliographystyle{abbrv}
\bibliography{bibfile}
%
%
%
%

\end{document}